\documentclass[a4paper,11pt]{article}
\usepackage{pos}
\usepackage{subcaption}
\usepackage{graphicx}
\usepackage{dsfont}
\usepackage{dcolumn}
\usepackage{multirow}
\usepackage{array}
\usepackage{url}
\usepackage{orcidlink}

\newcommand{\beq}{\begin{equation}}
\newcommand{\eeq}{\end{equation}}
\newcommand{\beqs}{\begin{eqnarray}}
\newcommand{\eeqs}{\end{eqnarray}}

\newcommand{\preprint}[1]{#1}

\newcommand{\orcidauthorBENNETT}{0000-0002-1678-6701}
\newcommand{\orcidauthorLUCINI}{0000-0001-8974-8266}
\newcommand{\orcidauthorPIAI}{0000-0002-2251-0111} 
\newcommand{\orcidauthorFORZANO}{0000-0003-0985-8858}
\newcommand{\orcidauthorVADACCHINO}{0000-0002-5783-5602}
\newcommand{\orcidauthorHONG}{0000-0002-3923-4184}
\newcommand{\orcidauthorLIN}{0000-0003-3743-0840}
\newcommand{\orcidauthorLEE}{0000-0002-4616-2422}
\newcommand{\orcidauthorZIERLER}{0000-0002-8670-4054}
\newcommand{\orcidauthorHSIAO}{0000-0002-8522-5190}
\renewcommand{\logo}{\relax}

\title{\boldmath Progress on the spectroscopy of an Sp(4) gauge theory coupled to matter in multiple representations}
\ShortTitle{Progress on the spectroscopy of Sp(4) gauge theory with matter in multiple representations}

\author*[a]{Ho Hsiao\,\orcidlink{\orcidauthorHSIAO}}
\author[b]{Ed Bennett\,\orcidlink{\orcidauthorBENNETT}}
\author[c]{Niccol\`o Forzano\,\orcidlink{\orcidauthorFORZANO}}
\author[d,e]{Deog Ki Hong\,\orcidlink{\orcidauthorHONG}}
\author[f]{Jong-Wan Lee\,\orcidlink{\orcidauthorLEE}}
\author[a,g]{C.-J. David Lin\,\orcidlink{\orcidauthorLIN}}
\author[b,h]{Biagio Lucini\,\orcidlink{\orcidauthorLUCINI}}
\author[c]{Maurizio Piai\,\orcidlink{\orcidauthorPIAI}}
\author[i]{Davide Vadacchino\,\orcidlink{\orcidauthorVADACCHINO}}
\author[c]{Fabian Zierler\,\orcidlink{\orcidauthorZIERLER}}

\affiliation[a]{Institute of Physics, National Yang Ming Chiao Tung University, Hsinchu, Taiwan}

\affiliation[b]{Swansea Academy of Advanced Computing, Swansea University, Swansea, United Kingdom}
\affiliation[c]{Department of Physics, Faculty of Science and Engineering, 
Swansea University, Singleton Park, Swansea, United Kingdom}
\affiliation[d]{Department of Physics, Pusan National University, Busan, Korea}
\affiliation[e]{Extreme Physics Institute, Pusan National University, Busan 46241, Korea}
\affiliation[f]{Particle Theory and Cosmology Group, Center for Theoretical Physics of the Universe, Institute for Basic Science (IBS), Daejeon, Korea}
\affiliation[g]{Centre for High Energy Physics, Chung-Yuan Christian University, Chung-Li, Taiwan}
\affiliation[h]{Department of Mathematics, Faculty of Science and Engineering, 
Swansea University, Singleton Park, Swansea, United Kingdom}
\affiliation[i]{Centre for Mathematical Sciences, University of Plymouth, Plymouth, United Kingdom}

\emailAdd{thepaulxiao.sc09@nycu.edu.tw}

\abstract{
We report progress on our lattice calculations for the mass spectra of low-lying composite states in the $Sp(4)$ gauge theory coupled to two and three flavors of Dirac fermions transforming in the fundamental and the two-index antisymmetric representations, respectively.
This theory provides an ultraviolet completion to the composite Higgs model with Goldstone modes in the $SU(4)/Sp(4)$ coset and with partial compositeness for generating the top-quark mass.
We measure the meson and chimera baryon masses.
These masses are crucial for constructing the composite Higgs model. In particular, the chimera baryon masses are important inputs for implementing top partial compositeness.
We employ Wilson fermions and the Wilson plaquette action in our simulations.
Techniques such as APE and Wuppertal smearing, as well as the procedure of generalised eigenvalue problem, are implemented in our analysis.
}

\FullConference{The 41st International Symposium on Lattice Field Theory (Lattice 2024)\\
July 28th - August 3rd, 2024\\
University of Liverpool\\}

\begin{document}
\makebox[0.95\textwidth][r]{\preprint{CTPU-PTC-24-35}}
\begin{center}
\includegraphics[width=\textwidth]{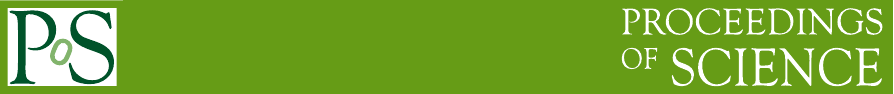}
\end{center}

\maketitle

\section{Introduction}

Interest in $Sp(2N)$ gauge theories, particularly in relation to composite Higgs models (CHMs)---see, for instance, the review in Ref.~\cite{Panico:2015jxa}---has motivated the development of the research programme of \textit{Theoretical Explorations on the Lattice with Orthogonal and Symplectic groups} (TELOS)~\cite{Bennett:2017kga, Bennett:2019jzz, Bennett:2019cxd, Bennett:2020hqd, Bennett:2020qtj, Bennett:2022yfa, Bennett:2022gdz, Bennett:2022ftz, Bennett:2023wjw, Bennett:2023gbe, Bennett:2023mhh, Bennett:2023qwx, Bennett:2024cqv, Bennett:2024wda}---see also Refs.~\cite{Kulkarni:2022bvh,Bennett:2023rsl, Dengler:2024maq,Bennett:2024bhy} for related works in the context of dark-matter physics.
To facilitate the construction of CHMs with top partial compositeness~\cite{Kaplan:1991dc}, it is essential to introduce fermions (hyperquarks) transforming in multiple representations. 
Since CHMs involve strongly coupled dynamics that necessitate non-perturbative treatment, lattice calculations naturally serve as a key tool for their investigation.
The existing literature on lattice gauge theories containing fermionic matter in multiple representations remains relatively sparse.
As a result, physicists have initiated efforts in this area and are conducting extensive explorations; see, e.g., Refs.~\cite{DeGrand:2016mxr, Ayyar:2017qdf, Ayyar:2017uqh, Ayyar:2017vsu, Ayyar:2018ppa, Ayyar:2018zuk, Cossu:2019hse, Lupo:2021nzv, DelDebbio:2022qgu, Golterman:2020pyx, Hasenfratz:2023sqa} for $SU(4)$, and Refs.~\cite{Bergner:2020mwl, Bergner:2021ivi} for $SU(2)$ gauge theories.

In this contribution, we discuss the $Sp(4)$ gauge theory coupled to $N_{\rm f}=2$ Dirac fermions in the fundamental, (f), and $N_{\rm as}=3$  Dirac fermions in the two-index antisymmetric, (as), representations of the gauge group.
This theory serves as the ultraviolet completion of one particular realisation~\cite{Barnard:2013zea} of a composite Higgs model (CHM) that provides a natural framework for accommodating a light Higgs boson~\cite{Kaplan:1983fs,
Georgi:1984af, Dugan:1984hq}---see also Refs.~\cite{Panico:2015jxa, 
Ferretti:2013kya,Ferretti:2016upr,Cacciapaglia:2019bqz, Cacciapaglia:2020kgq}.
The global symmetry breaking pattern in the fundamental-hyperquark sector of this theory is $SU(4)/Sp(4)$, leading to the emergence of five PNGBs.
Four of these PNGBs can be interpreted as the Standard Model (SM) Higgs doublet, leaving only one additional Goldstone mode.
The reality of the (as) representation results in the global symmetry-breaking pattern $SU(6)/SO(6)$.
The $SU(3)$ subgroup within the unbroken $SO(6)$ can be gauged and interpreted as the QCD colour group~\cite{Barnard:2013zea, Ferretti:2013kya}.
This theory includes two families of mesons, each consisting of a pair of valence hyperquarks and anti-hyperquarks, both in either the (f) or (as) sector of the theory.
We refer to these as fundamental and antisymmetric mesons, respectively.
In addition, the theory admits exotic fermionic bound states, known as chimera baryons, composed of two (f) and one (as) hyperquarks, combined to form a hypercolour-singlet of the $Sp(4)$ gauge group.
The existence of these objects enables the implementation of partial compositeness for the top quark~\cite{Kaplan:1991dc}, wherein chimera baryons, engineered to share the same quantum numbers as the top quark, can generate the mass of the latter through mixing.
In the following, we review earlier work on the spectrum~\cite{Bennett:2017kga, Bennett:2019jzz, Bennett:2019cxd, Bennett:2022yfa, Bennett:2023mhh, Bennett:2024wda} and present preliminary new results on work currently in progress, which will form the foundation for our future publications.

\section{Lattice field theory, observables, preliminary tests.}

In order to study the strongly-coupled $Sp(4)$ gauge theories, we employ lattice field theory and discretise the four-dimensional Euclidean space-time.
Gauge field dynamics are described by the standard Wilson plaquette action, and the hyperquarks are represented by Wilson fermions in the Hybrid Monte Carlo (HMC)~\cite{Duane:1987de} and Rational HMC (RHMC)~\cite{Clark:2006fx} simulations.
We adopt the gradient flow method~\cite{Luscher:2010iy, Luscher:2011bx, Luscher:2013vga} as a scale-setting procedure, expressing all masses in terms of the gradient flow scale, $w_0$~\cite{BMW:2012hcm}, using the notation $\hat{m}=w_0m$.
Our lattice simulations and measurements are conducted by using two software packages, \textit{HiRep}~\cite{DelDebbio:2008zf} and \textit{GRID}~\cite{Boyle:2015tjk}, with the add-ons for the $Sp(2N)$ gauge group~\cite{Bennett:2017kga, Bennett:2023gbe}.
Table~\ref{tab:ensembles} summarises the main properties of the ensembles generated for the fully dynamical calculations,  with results measured on these ensembles to be discussed later.

\begin{table}
    \centering
    \caption{Ensembles used in the fully dynamical calculations. The inverse coupling is denoted by $\beta$, and the bare masses of the two fermion species by $am^{\rm f}_0$ and $am^{\rm as}_0$, where $a$ is the lattice spacing. The number of sites in the temporal and spatial directions are given by $N_t$ and $N_s$, respectively.
    The number of thermalisation steps excluded from the analysis is denoted as $N_{\rm therm}$, while $n_{\rm skip}$ indicates the number of trajectories discarded between retained configurations. 
    The number of configurations used in the analysis is $N_{\rm conf}$. 
    The average plaquette value of each ensemble is denoted by $\langle P \rangle$, and $w_0 / a$ represents the Wilson flow scale.}
    \small
    \begin{tabular}{|c|c|c|c|c|c|c|c|c|c|c|c|c|}
    
    \hline \hline
    Label & $\beta$ & $am_0^{\rm as}$ & $am_0^{\rm f}$ & $N_t$ & $N_s$ & $N_{\rm therm}$ & $n_{\rm skip}$ & $N_{\rm conf}$ & $\langle P \rangle$ & $w_0 / a$  \\ 
    \hline
        M1 & 6.5 & -1.01 & -0.71 & 48 & 20 & 3006 & 14 & 479 & 0.585172(16) & 2.5200(50) \\ 
        M2 & 6.5 & -1.01 & -0.71 & 64 & 20 & 1000 & 28 & 698 & 0.585172(12) & 2.5300(40) \\ 
        M3 & 6.5 & -1.01 & -0.71 & 96 & 20 & 4000 & 26 & 436 & 0.585156(13) & 2.5170(40) \\ 
        M4 & 6.5 & -1.01 & -0.70 & 64 & 20 & 1000 & 20 & 709 & 0.584228(12) & 2.3557(31) \\ 
        M5 & 6.5 & -1.01 & -0.72 & 64 & 32 & 3020 & 20 & 295 & 0.5860810(93) & 2.6927(31) \\ 
    \hline \hline
    \end{tabular}
    \label{tab:ensembles}
\end{table}

The mass spectra are extracted from measurements of two-point correlation functions involving the relevant meson and chimera baryon operators. 
All meson interpolating operators, including pseudoscalar, vector, tensor, axial-vector, axial-tensor, and scalar mesons composed of either fundamental or antisymmetric hyperquarks, are listed in an earlier publication~\cite{Bennett:2017kga}.
These mesons are denoted as PS (ps), V (v), T (t), AV (av), AT (at), and S (s), respectively, with fundamental and antisymmetric mesons distinguished by uppercase and lowercase subscripts, respectively.
The chimera-baryon correlators use operators of the following form:
 \beqs
 {\mathcal{O}}_{\rho}^{ijk,5} &\equiv& Q^{i\,a}_{\alpha} (C\gamma^{5})_{\alpha\beta} Q^{j\,b}_{\beta} \Omega^{ad}\Omega^{bc} \Psi^{k\,cd}_{\rho} \, , \label{eq:chim_bar_src} \\ 
 {\mathcal{O}}_{\rho}^{ijk,\mu} &\equiv& Q^{i\,a}_{\alpha} (C\gamma^{\mu})_{\alpha\beta} Q^{j\,b}_{\,\beta} \Omega^{ad}\Omega^{bc} \Psi^{k\,cd}_{\rho} \, ,\label{eq:chim_bar_src_mu}
 \eeqs
where $Q$ and $\Psi$ are (f) and (as) hyperquarks, respectively; $a,\,b,\,c,\,d$ are hypercolour indices; $\alpha, \,\beta,\, \rho$ are spinor indices; $i,\,j$ are (f)-type flavor indices, $k$ is an (as)-type flavor index; $\gamma^5$ and $\gamma^{\mu}$ are $4\times 4$ Dirac matrices; $C$ is the charge conjugation matrix; and $\Omega$ is the symplectic $4\times 4$ matrix.
Since both operators in Eqs.~(\ref{eq:chim_bar_src}) and~(\ref{eq:chim_bar_src_mu}) overlap with parity-even and parity-odd states, and the operator ${\mathcal{O}}^{\mu}$ couples to both spin-1/2 and spin-3/2 states, we apply parity and spin projections to isolate the states with the desired quantum numbers---see details in Section III.A of Ref.~\cite{Bennett:2023mhh}.
The lightest state sourced by the operator ${\mathcal{O}}^{5}$, labeled as $\Lambda_{\rm CB}$, is a spin-1/2 state.
For ${\mathcal{O}}^{\mu}$, the lightest state with spin-1/2 and spin-3/2 are denoted as $\Sigma_{\rm CB}$ and $\Sigma^\ast_{\rm CB}$, respectively.
Both $\Lambda_{\rm CB}$ and $\Sigma_{\rm CB}$ are top-partner candidates~\cite{Gripaios:2009pe, Banerjee:2022izw}.

\begin{figure}
	\centering
	\includegraphics[width=0.80\textwidth]{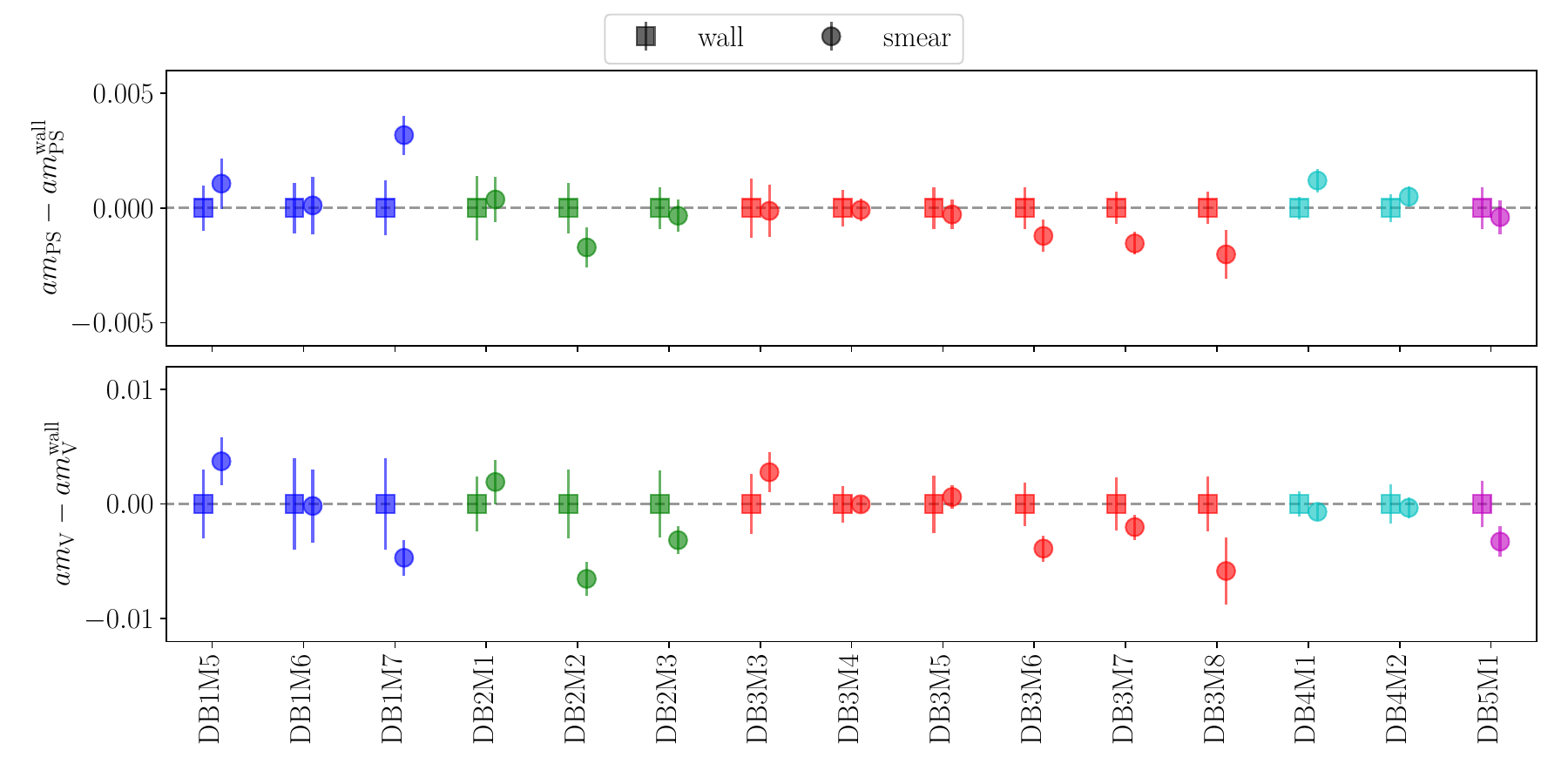}
	\caption{
    Comparison of pseudoscalar (upper panel) and vector (lower panel) meson masses composed of (f) hyperquarks, extracted from correlation functions using wall sources and smearing techniques.
    The measurements are performed on ensembles with (f)-type dynamical fermions taken from Ref.~\cite{Bennett:2019jzz}, from which we borrow the conventional naming of the individual ensembles (the horizontal axis).
    Data points from the same ensemble are slightly shifted horizontally to distinguish between two types of techniques.
    The lattice couplings used are $\beta= 6.9$ (blue), $7.05$ (green), $7.2$ (red), $7.4$ (cyan), and $7.5$ (magenta).
    }
	\label{fig:smearing}
\end{figure}

\begin{figure}
	\centering
	\includegraphics[width=0.48\textwidth]{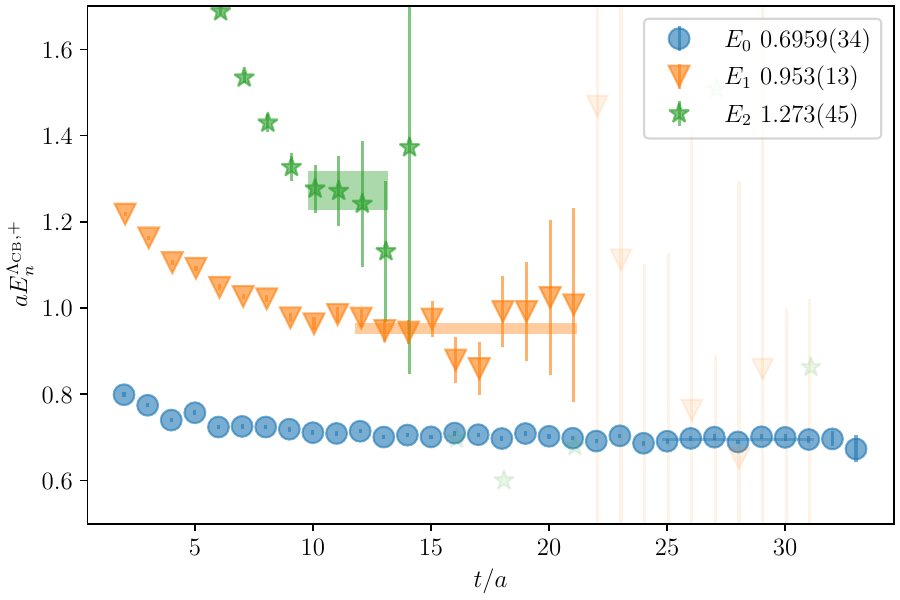}
    \includegraphics[width=0.48\textwidth]{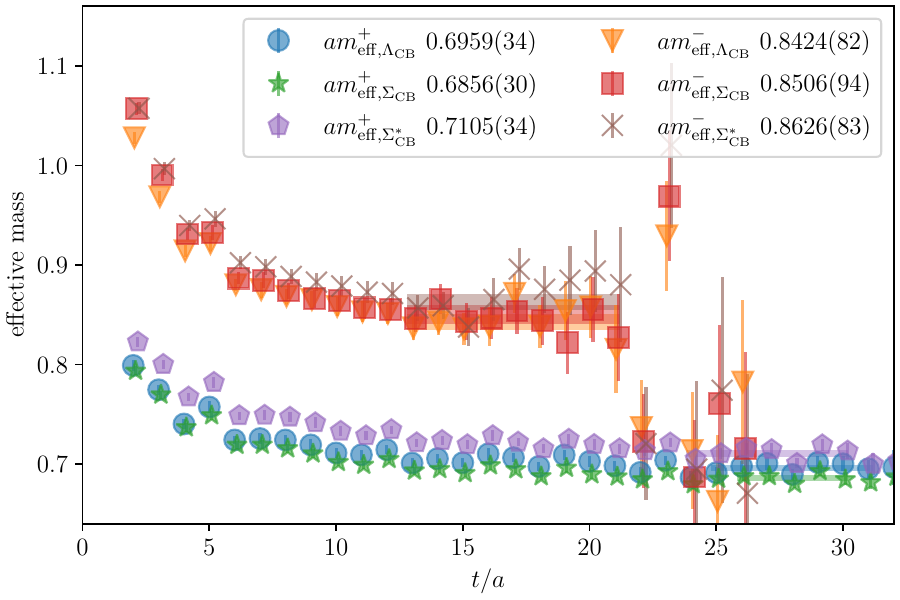}
	\caption{Effective mass plot of chimera baryon masses, measured in the ensemble M5 listed in Table~\ref{tab:ensembles}. Left panel:  energy states of the parity-even $\Lambda_{\rm CB}$, obtained by solving GEVP with three different levels of smearing at the source and sink. Right panel:  ground-state effective mass for all available chimera baryons. 
    }
	\label{fig:CB_spec}
\end{figure}

In addition to using stochastic wall sources~\cite{Boyle:2008rh} for two-point function measurements, our analysis incorporates results obtained through the implementation of Wuppertal~\cite{Gusken:1989qx} and APE smearings~\cite{APE:1987ehd} as signal optimisation techniques.
Figure~\ref{fig:smearing} compares the fundamental pseudoscalar and vector meson masses calculated using wall sources and smearing techniques on the ensembles generated for the study in Ref.~\cite{Bennett:2019jzz}.
We confirm that the application of smearing techniques not only reduces uncertainty but also produces measurements compatible with those obtained using wall sources within the 2$\sigma$ interval.
To further enhance the effectiveness of our measurement strategy, we reformulate it as a generalised eigenvalue problem (GEVP) and incorporate a basis of operators constructed with varying levels of smearing.
Figure~\ref{fig:CB_spec} illustrates, in lattice units, the results obtained through this analysis procedure for computing the mass of the chimera baryons.
The left panel shows the energy states of the parity-even $\Lambda_{\rm CB}$, denoted as $aE_{n}^{\Lambda_{\rm CB}, +}$.
Similarly, the energy states of the $\Sigma_{\rm CB}$ and $\Sigma_{\rm CB}^\ast$ for both parities are determined using the same methodology.
In the right panel, the effective masses of each chimera baryon ground state are displayed, demonstrating that parity-even states are lighter than their parity-odd counterparts.

\section{Mass spectra}

\begin{figure}
	\centering
	\includegraphics[width=0.80\textwidth]{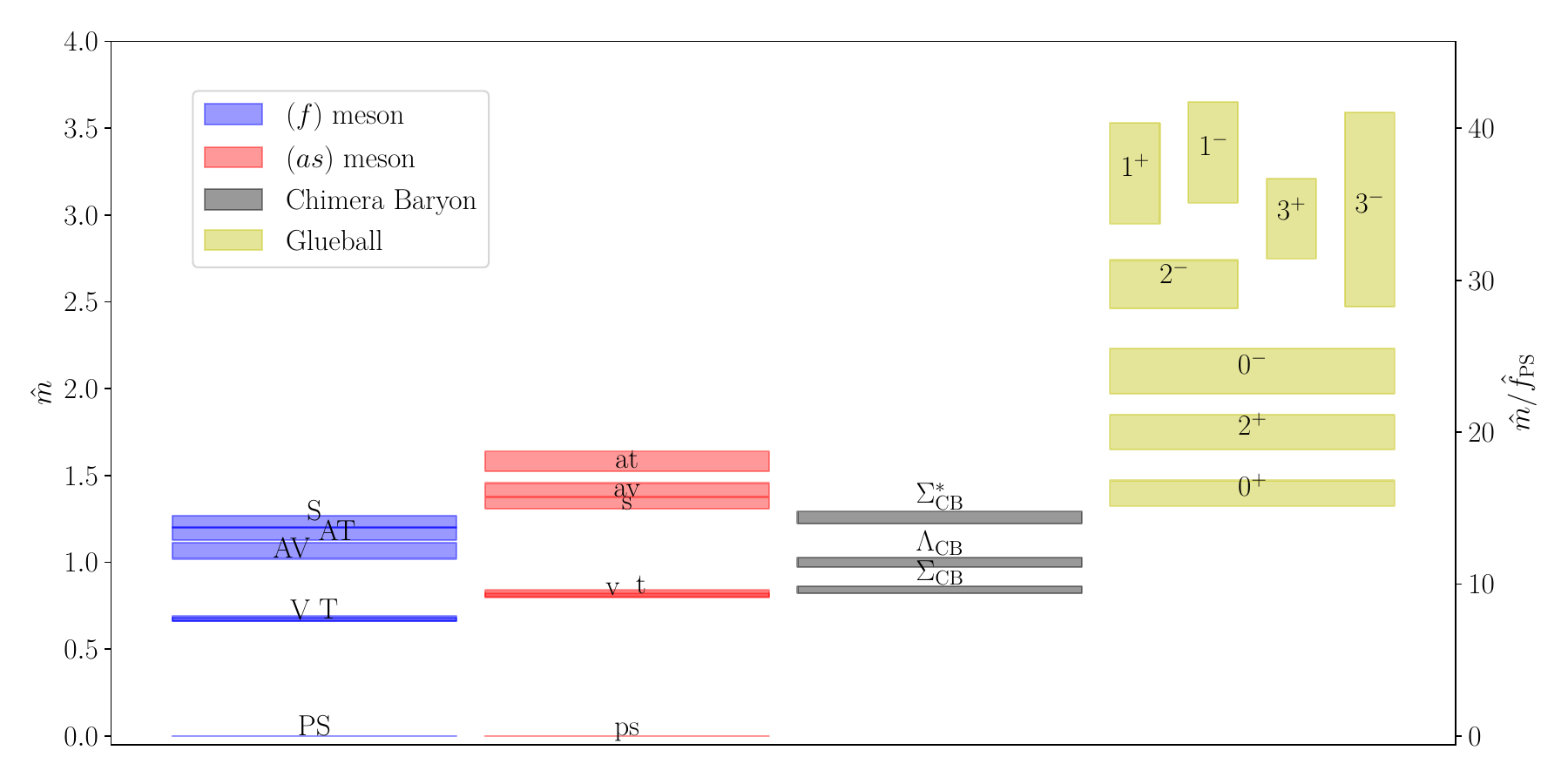}
	\caption{The mass spectrum in the massless and continuum limits of the $Sp(4)$ gauge theory computed in the quenched approximation. Pseudoscalar, vector, tensor, axial-vector, axial-tensor and scalar mesons composed of fundamental (antisymmetric)  hyperquarks are denoted as PS (ps), V (v), T(t), AV (av), AT (at) and S (s), respectively. We denote as
    $\Lambda_{\rm CB}$, $\Sigma_{\rm CB}$, and $\Sigma_{\rm CB}^{\ast}$ the lightest, parity-even chimera baryons.
    Glueball states are labelled by their spin and parity quantum numbers, $J^{P}$.
    This figure is taken from Ref.~\cite{Bennett:2023mhh}.
    }
	\label{fig:quench_spec}
\end{figure}

\begin{figure}
	\centering
	\includegraphics[width=0.85\textwidth]{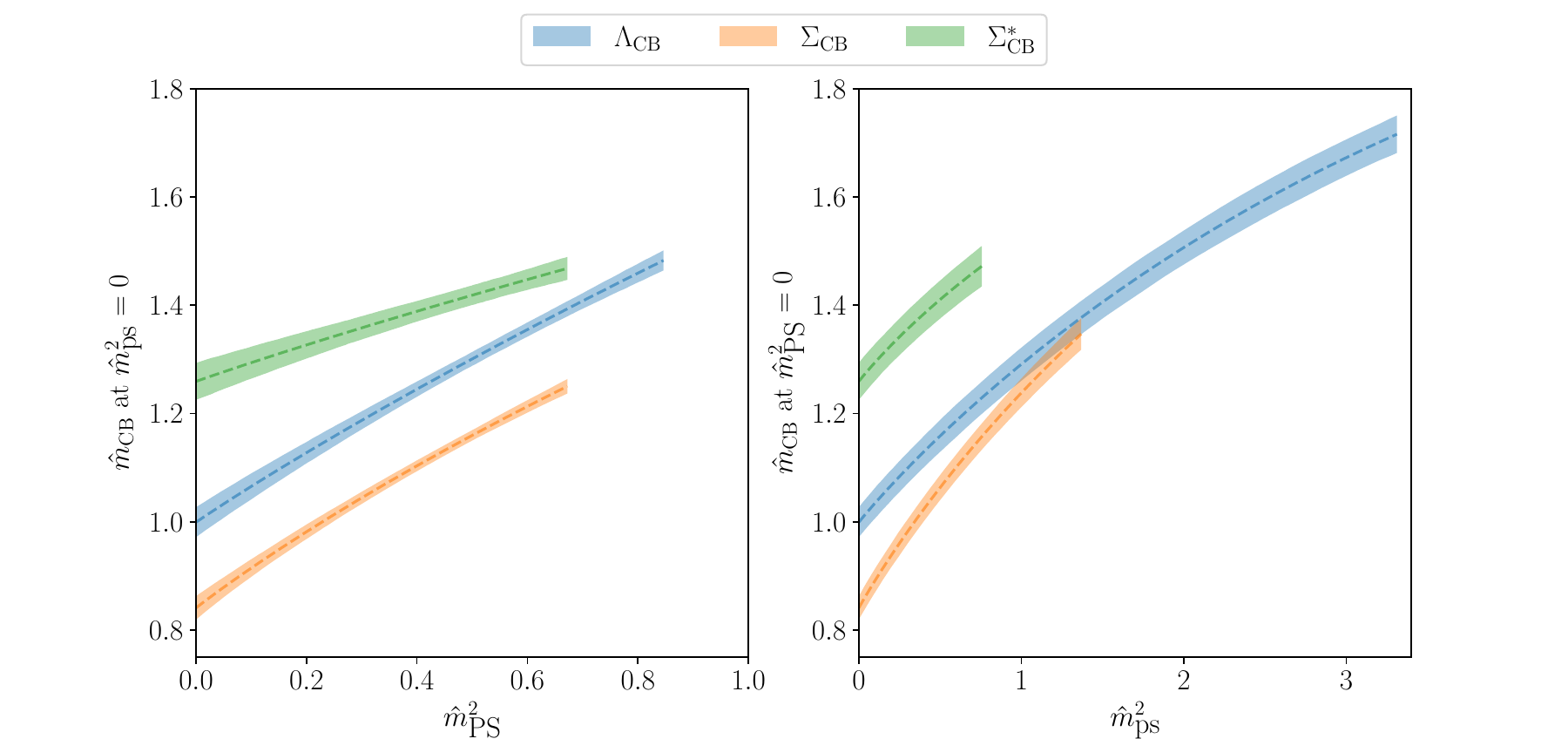}
	\caption{Mass dependence of the chimera baryons, in the continuum limit, as a function of the mass squared of the pseudoscalars, $\hat{m}^{2}_{\rm PS}$ (left) and $\hat{m}^{2}_{\rm ps}$ (right). This figure is taken from Ref.~\cite{Bennett:2023mhh}.
    }
	\label{fig:CB_extrapolation}
\end{figure}

The TELOS collaboration is the first to conduct systematic, dedicated lattice studies on the spectroscopy of $Sp(4)$ gauge theories.
This summary begins with the case where only the gauge dynamics are captured in the Monte Carlo simulations, while fermions are treated in the quenched approximation.
On the one hand, this initial series of measurements enables us to test the necessary measurement techniques in a resource-efficient way and establish benchmarks for future calculations with dynamical fermions.
On the other hand, these results provide a reasonable approximation of the spectra in the physical regime where fermions are not particularly light, which is of phenomenological interest in the contexts of CHMs and dark matter.
The mass of the two families of flavored mesons, composed of (f) and (as) hyperquarks~\cite{Bennett:2019cxd}, along with those of glueballs~\cite{Bennett:2020hqd, Bennett:2020qtj}, and our first spectrum measurement of chimera baryons~\cite{Bennett:2023mhh}, are summarised in Figure~\ref{fig:quench_spec}.
We report only parity-even chimera baryons, which are lighter than their parity partners.
The results are presented in units of both the gradient flow scale, $w_0$, and the decay constant of the pseudoscalars made of (f) hyperquarks, $f_{\rm PS}$, extracted from the measurement of the relevant correlation functions~\cite{Bennett:2019cxd} and renormalised at the one-loop level with tadpole improvement~\cite{Martinelli:1982mw, Lepage:1992xa}.
The masses displayed are obtained by extrapolating the measurements toward continuum and massless limits,
using a simplified approach inspired by Wilson~\cite{Sheikholeslami:1985ij, Rupak:2002sm} and heavy-baryon~\cite{Jenkins:1990jv, Bernard:1995dp, Beane:2003xv} chiral perturbation theory. 
The Akaike Information Criterion (AIC)~\cite{Akaike:1998zah, Jay:2020jkz} is used to evaluate the fit quality and identify the optimal fitting procedure for each chimera baryon in such extrapolations.
Figure~\ref{fig:CB_extrapolation} presents the dependence of chimera baryon masses on pseudoscalar masses in the continuum limit.
We observe a consistent mass hierarchy, except in the heavy (as) pseudoscalar meson mass region, where $\Lambda_{\rm CB}$ and $\Sigma_{\rm CB}$ become degenerate.

\begin{figure}
	\centering
	\includegraphics[width=0.31\textwidth]{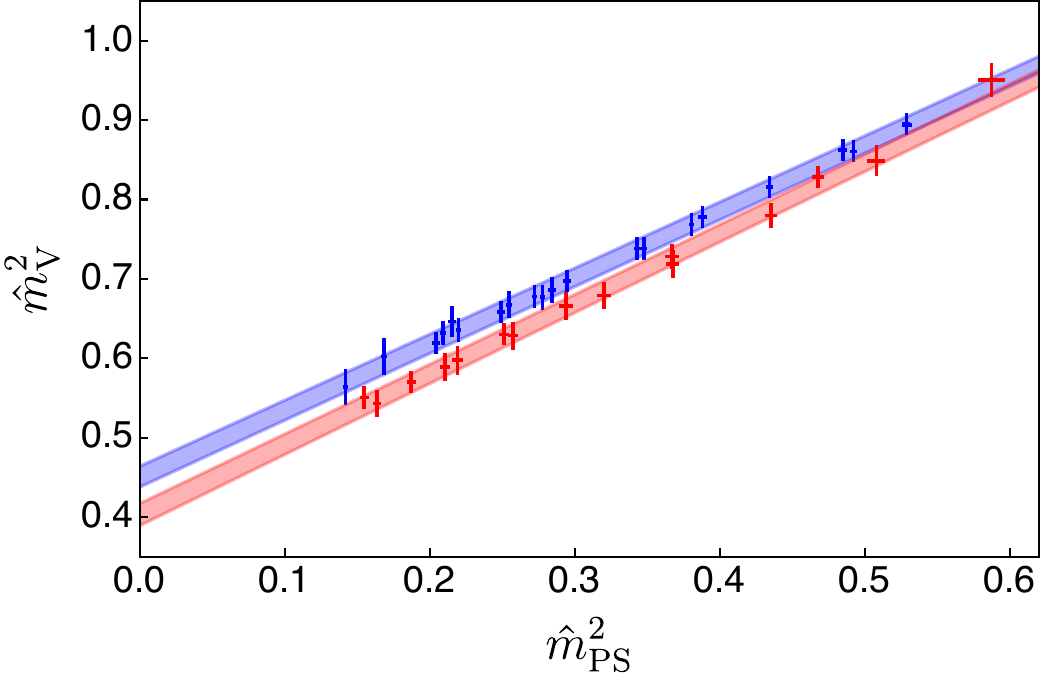}
    \includegraphics[width=0.31\textwidth]{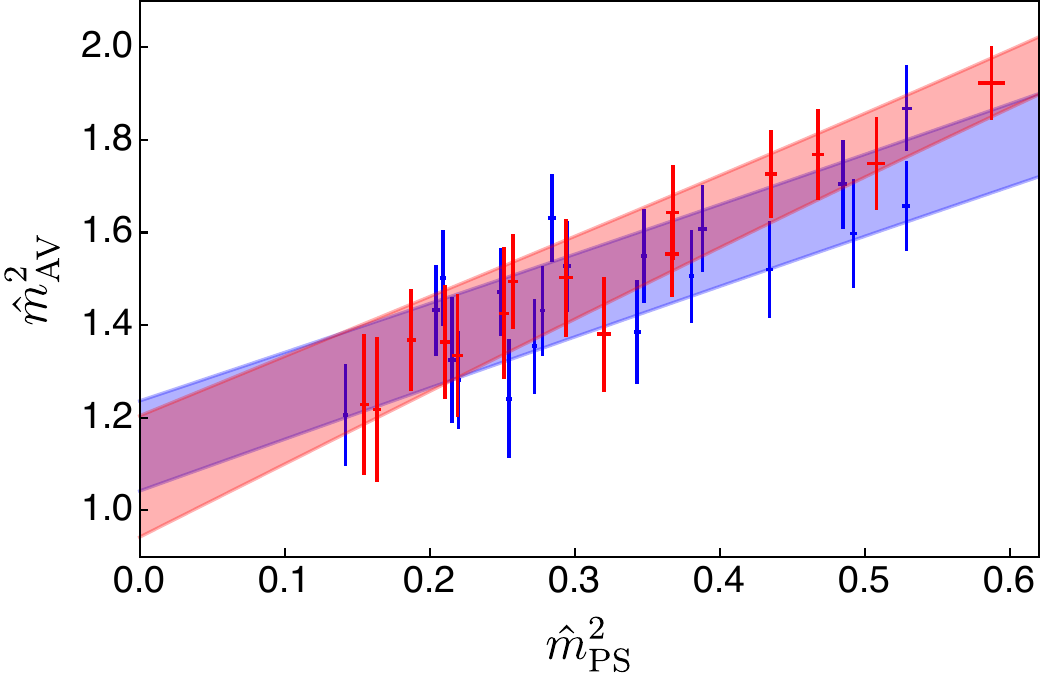}
    \includegraphics[width=0.31\textwidth]{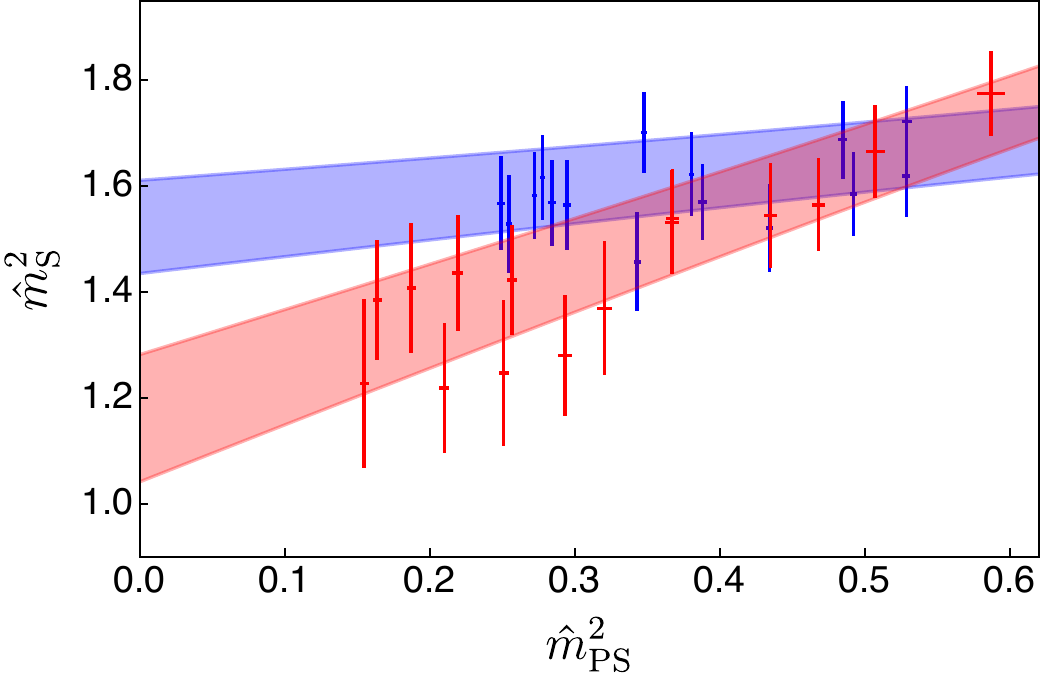}
	\caption{
    Comparison of quenched (blue) and dynamical (red) calculations for the extrapolations of the squared masses of flavored vector (V), axial-vector (AV), and scalar (S) mesons composed of (f)-type hyperquarks, plotted as a function of the (f) pseudoscalar (PS) meson mass squared.
    These plots are taken from Ref.~\cite{Bennett:2019jzz}.
    }
	\label{fig:quench_effects}
\end{figure}

The natural next step is to introduce dynamical hyperquarks of one species, transforming either in the fundamental representation~\cite{Bennett:2019jzz} or antisymmetric representation~\cite{AS}---while quenching the other species of hyperquarks.
In Figure~\ref{fig:quench_effects}, we compare the masses of mesons composed of (f) hyperquarks computed in the quenched approximation with those in the dynamical case to quantify the size of quenching effects.
In the region of parameter space where the measurements are performed, the discrepancies are modest. Even in the extrapolations toward the combined continuum and massless limits, they remain approximately $\sim$$10\%$ for the vector mesons and $\sim$$25\%$ for the scalar mesons.
The axial-vector mesons exhibit smaller discrepancies, though they are accompanied by larger uncertainties.
Additionally, the first measurement of flavor-singlet mesons in the theory with dynamical (f) hyperquarks is reported in Ref.~\cite{Bennett:2023rsl}.

\begin{figure}
	\centering
	\includegraphics[width=0.80\textwidth]{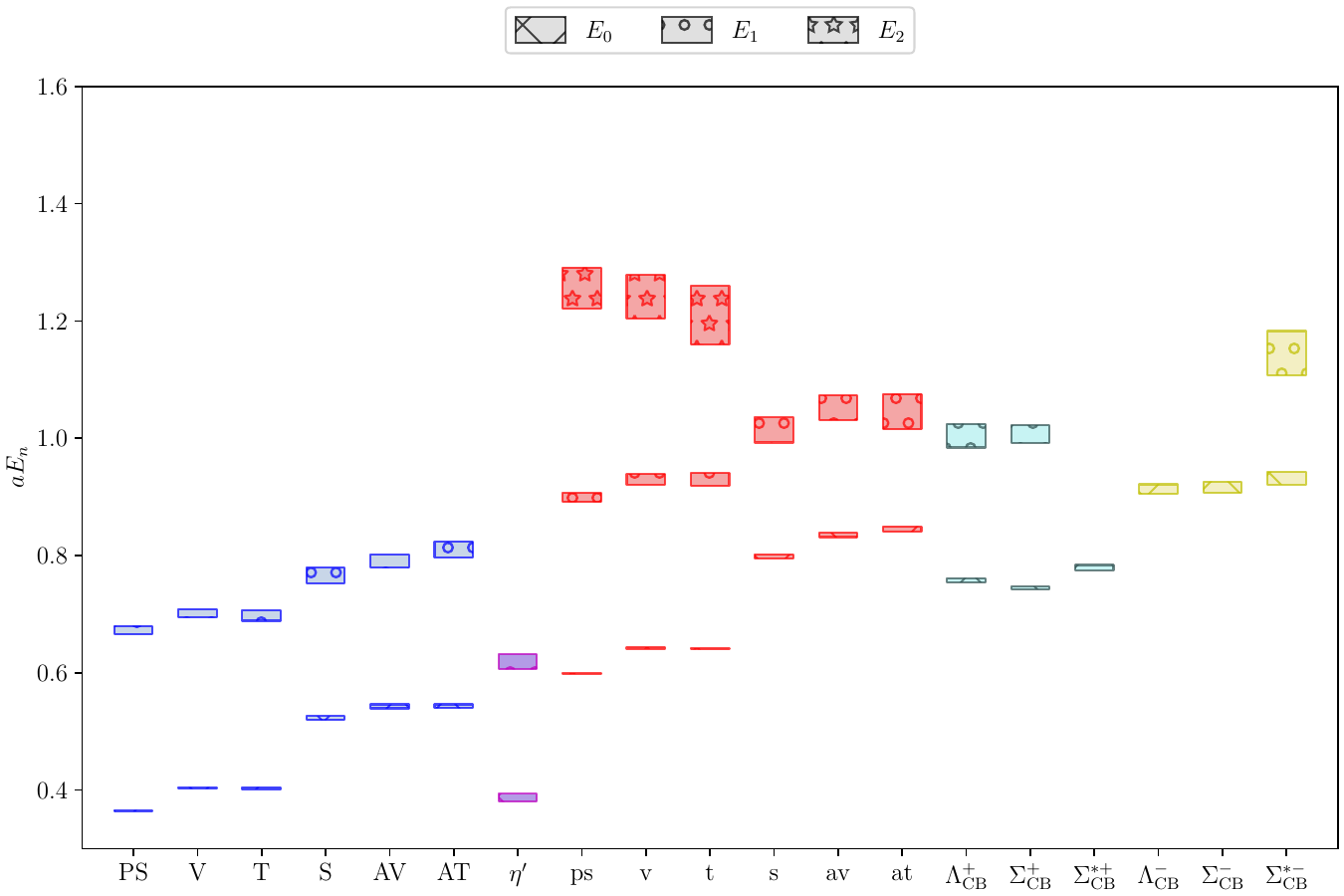}
	\caption{The mass spectrum  of all accessible flavored and flavor-singlet mesons, and of chimera baryons of both parities, including, where possible, also excited states, measured in 
	 ensemble M2 in Table~\ref{tab:ensembles}, obtained by applying the GEVP method with smeared correlation functions.
    }
	\label{fig:dyn_full_spec_M2}
\end{figure}

\begin{figure}[t]
	\centering
	\includegraphics[width=0.80\textwidth]{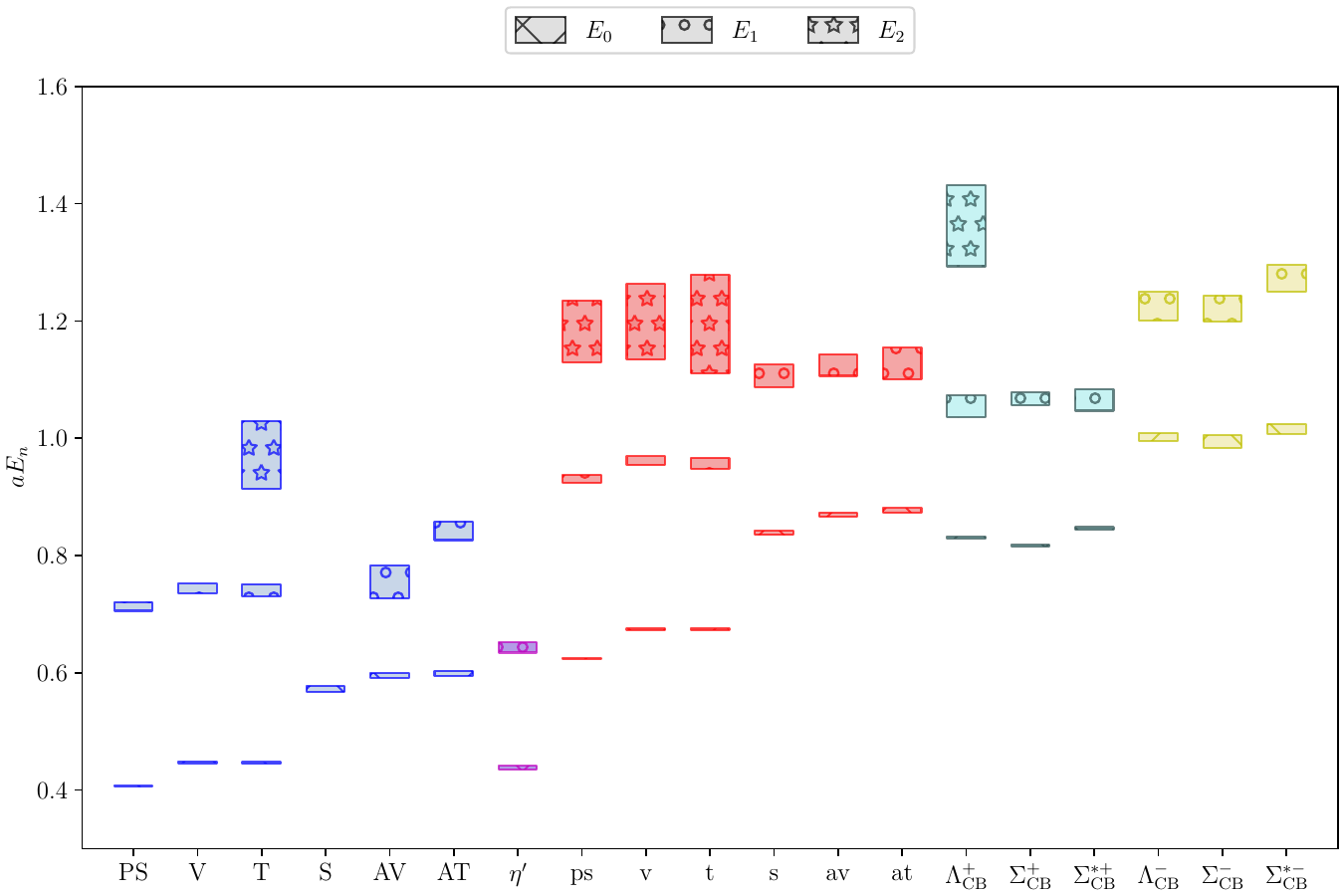}
	\caption{The mass spectrum  of all accessible flavored and flavor-singlet mesons, and of chimera baryons of both parities, including, where possible, also excited states, measured in 
	 ensemble M4 in Table~\ref{tab:ensembles}, obtained by applying the GEVP method with smeared correlation functions.
    }
	\label{fig:dyn_full_spec_M4}
\end{figure}

\begin{figure}[t]
	\centering
	\includegraphics[width=0.80\textwidth]{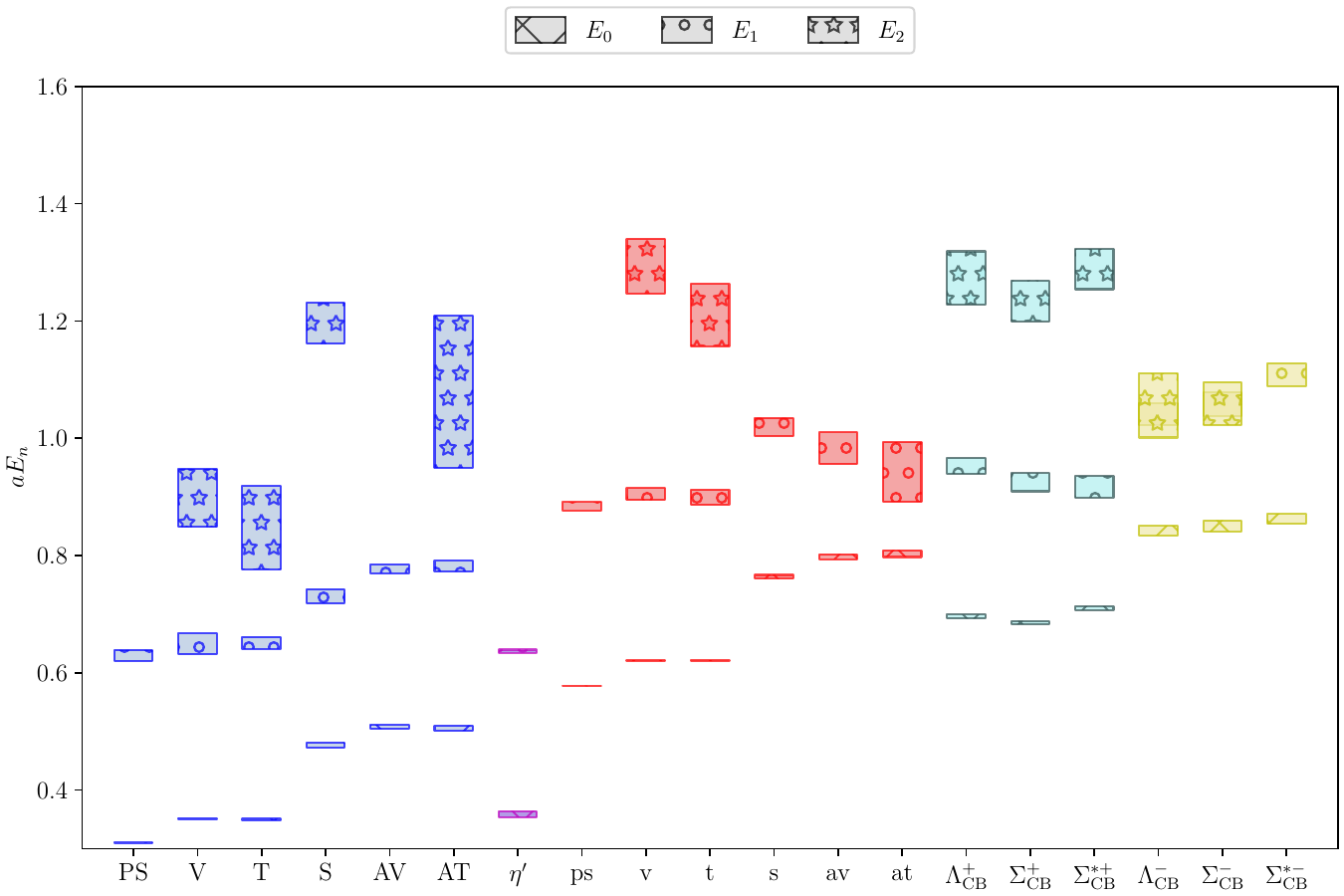}
	\caption{The mass spectrum  of all accessible flavored and flavor-singlet mesons, and of chimera baryons of both parities, including, where possible, also excited states, measured in 
	 ensemble M5 in Table~\ref{tab:ensembles}, obtained by applying the GEVP method with smeared correlation functions.
    }
	\label{fig:dyn_full_spec_M5}
\end{figure}

The study of the dynamical theory with hyperquarks in multiple representations begins with an exploration of the lattice parameter space~\cite{Bennett:2022yfa}.
For spectroscopy studies, we generate five ensembles, as detailed in Table~\ref{tab:ensembles}.
These ensembles serve as a testing ground for the spectral density method discussed in Ref.~\cite{Bennett:2024cqv}, 
applied to the mesonic two-point functions.
We also conduct the first study of the mixing effects between flavor-singlet mesons involving hyperquarks in both representations in Ref.~\cite{Bennett:2024wda}.
Furthermore, we present preliminary results for chimera baryon masses, covering both parity channels and excited states in Figs.~\ref{fig:dyn_full_spec_M2},~\ref{fig:dyn_full_spec_M4}, and~\ref{fig:dyn_full_spec_M5}, measured on ensembles M2, M4, and M5 listed in Table~\ref{tab:ensembles}, respectively.
These three figures display a combination of spectra of accessible states, including the pseudoscalar, vector, tensor, axial-vector, axial-tensor and scalar flavored mesons composed of fundamental (antisymmetric) hyperquarks---denoted as PS (ps), V (v), T(t), AV (av), AT (at) and S (s), respectively---the pseudoscalar flavor-singlet meson, labelled as $\eta^\prime$ obtained from the mixing of the two meson sectors, and chimera baryons, $\Lambda_{\rm CB}$, $\Sigma_{\rm CB}$, and $\Sigma^\ast_{\rm CB}$, with both even (+) and odd (-) parities. Where possible, we also show excited states.
A complete analysis of the spectroscopy will be prepared for publication in the near future.

\section{Summary and Outlook}

We have reviewed results on the spectrum of bound states in the $Sp(4)$ gauge theory coupled to $N_{\rm f}=2$ Dirac fermions transforming in the fundamental and $N_{\rm as}=3$ Dirac fermions transforming in the two-index antisymmetric representations, simulated in both the quenched approximation and partially dynamical lattice setups.
Building on the experience gained from these studies, we presented preliminary results obtained with the fully dynamical theory.
The mass spectra shown here demonstrate that, in the available region of parameter space, the combination of Wuppertal smearing, APE smearing, and an optimised GEVP analysis allows access to a wide variety of bound states.
These include mesons with different quantum numbers, chimera baryons of both parities, and their excited states.

The next steps of this ambitious programme will involve combining these measurements with off-shell observables extracted using the spectral density methodology applied to meson and chimera baryon correlation functions. 
We also envision the possibility of reanalysing the whole spectrum of flavor-singlet states, incorporating not only meson but also glueball observables into the GEVP framework.
In the long run, we aim to compute other non-trivial off-shell observables, particularly matrix elements that are relevant in the contexts of CHM and partial compositeness.
Achieving these advanced objectives will require a significant upgrade to the technology used for ensemble generation, by implementing improvements and potentially adopting a different type of fermion in simulations to more effectively approach the continuum and massless limits.

\begin{acknowledgments}

{\small

EB and BL are supported by the EPSRC ExCALIBUR programme ExaTEPP (project EP/ X017168/1).  
EB, BL, MP, FZ are supported by the STFC Consolidated Grant No.~ST/X000648/1. 
EB is supported by the STFC Research Software Engineering Fellowship EP/V052489/1.
NF is supported by the STFC Consolidated Grant No.~ST/X508834/1. 
DKH is supported by Basic Science Research Program through the National Research Foundation of Korea (NRF) funded by the Ministry of Education (NRF-2017R1D1A1B06033701) and the NRF grant MSIT 2021R1A4A5031460 funded by the Korean government.
JWL is supported by IBS under the project code IBS-R018-D1. 
HH and CJDL are supported by the  Taiwanese MoST grant 109-2112-M-009-006-MY3 and NSTC grant 112-2112-M-A49-021-MY3.
CJDL is also supported by Grants No.~112-2639-M-002-006-ASP and No.~113-2119-M-007-013. 
BL and MP have been supported by the STFC Consolidated Grant No.~ST/T000813/1 and by the European Research Council (ERC) under the European Union’s Horizon 2020 research and innovation program under Grant Agreement No.~813942. 
DV is supported by the STFC under Consolidated Grant No.~ST/X000680/1.

Numerical simulations have been performed on the DiRAC Extreme Scaling service at The University of Edinburgh, and on the DiRAC Data Intensive service at Leicester. 
The DiRAC Extreme Scaling service is operated by the Edinburgh Parallel Computing Centre on behalf of the STFC DiRAC HPC Facility (www.dirac.ac.uk). 
This equipment was funded by BEIS capital funding via STFC capital grant ST/R00238X/1 and STFC DiRAC Operations grant ST/R001006/1.
DiRAC is part of the UKRI Digital Research Infrastructure.

{\bf Open Access Statement}---For the purpose of open access, the authors have applied a Creative Commons 
Attribution (CC BY) licence to any Author Accepted Manuscript version arising.

{\bf Research Data Access Statement}---The results reported here are based on preliminary analysis.
Further analysis and the data generated for this manuscript will be released together with an upcoming publication. Alternatively, preliminary data and code can be obtained from the authors upon request.

} 

\end{acknowledgments}


{\footnotesize
\bibliographystyle{JHEP}
\bibliography{bibliography.bib}
}

\end{document}